\documentclass{ws-procs10x7}
\usepackage{balance}
\usepackage{epsfig}

\def\ta{\theta_{12}}
\def\tb{\theta_{23}}
\def\tc{\theta_{13}}
\def\Dsol{\Delta_{21}}
\def\Datm{\Delta_{32}}

\def\be{\begin{equation}}
\def\ee{\end{equation}}
\def\bea{\begin{eqnarray}}
\def\eea{\end{eqnarray}}
\def\bes{\begin{subequations}}
\def\ees{\end{subequations}}
\def\ba{\begin{array}}
\def\ea{\end{array}}
\def\beq{\begin{equation}}
\def\eeq{\end{equation}}

\def\barr{\begin{eqnarray}}
\def\earr{\end{eqnarray}}

\def\dmsq{\Delta m^2}
 
\def\lsim{\mathrel{\rlap{\lower4pt\hbox{\hskip1pt$\sim$}}
    \raise1pt\hbox{$<$}}}                
\def\gsim{\mathrel{\rlap{\lower4pt\hbox{\hskip1pt$\sim$}}
    \raise1pt\hbox{$>$}}}                

\newcommand{\ch}{{\chi^2}}

\newcolumntype{d}[1]{D{.}{.}{#1}}

\makeindex
\begin{document}

\title{Constraints on flavor-dependent long range forces from
neutrino experiments}

\author{Abhijit Bandyopadhyay \and Amol Dighe$^*$}

\address{Tata Institute of Fundamental Research, Mumbai 400 005, India\\
$^*$E-mail: amol@theory.tifr.res.in}

\author{Anjan S. Joshipura}

\address{Physical Research Laboratory, Ahmedabad 380 009, India }


\twocolumn[\maketitle\abstract{
We study the impact of 
flavor-dependent long range leptonic forces
mediated by the $L_e-L_\mu$ or $L_e -L_\tau$ gauge bosons
on the solar neutrino oscillations,
when the interaction range $R_{LR}$ is much larger than the
Earth-Sun distance.
The solar and atmospheric neutrino mass scales do not get 
trivially decoupled in this situation
even if $\theta_{13}$ is vanishingly small.
In addition, for $\alpha \gsim 10^{-52}$ and normal hierarchy,
resonant enhancement of $\theta_{13}$ may give rise to strong 
energy dependent effects on the $\nu_e$ survival probability.
A complete three generation analysis of the 
solar neutrino and KamLAND data gives 
the $3\sigma$ limits
$\alpha_{e\mu} < 3.4 \times 10^{-53}$ and
$\alpha_{e\tau} < 2.5 \times 10^{-53}$
when $R_{LR}$ is
much smaller than our distance from the galactic center.
With larger $R_{LR}$,
the collective LR potential due to all the electrons in the
galaxy becomes significant and the constraints on $\alpha$
become stronger by upto two orders of magnitude.
}

] 

\section{Introduction}
\label{intro}

Flavor-dependent long range (LR) leptonic forces,
like those mediated by the $L_e-L_\mu$ or $L_e -L_\tau$ 
gauge bosons,\cite{masso}
constitute a minimal extension of the
standard model that preserves its renormalizability.
The extra $U(1)$ gauge boson, albeit nearly massless,
would escape direct detection if it couples 
to the matter very weakly.\cite{mj1}
Since these long range forces violate the equivalence principle,
they are strongly constrained by the lunar ranging
and E\"otv\"os type gravity experiments.\cite{lr1,lr2}
For a range of the Earth-Sun distance or more, 
these experiments imply the 2$\sigma$ bounds 
$\alpha < 3.4 \times 10^{-49}$, 
where $\alpha$
denotes the strength of the long range potential.

The coupling of the
solar electrons to the $L_e-L_\beta$ gauge bosons 
generates a long range potential $V_{e\beta}^\odot$ for neutrinos, 
whose value
at the Earth is $1.3 \times 10^{-11} {\rm eV} (\alpha_{e\beta}/10^{-50})$.
The typical value of $\dmsq/E$ for atmospheric neutrinos is
$\sim 10^{-12}$ eV, so even with the strong constraints above,
LR forces affect atmospheric neutrino oscillations.
This allows one to put stronger constraints\cite{mj1} 
on the couplings,
$\alpha_{e\mu} < 5.5 \times 10^{-52}$ and 
$\alpha_{e\tau} < 6.4 \times 10^{-52}$.

\section{LR potential due to the Sun, the Earth and the galaxy}
\label{formalism}

\begin{figure}
\centerline{\psfig{figure=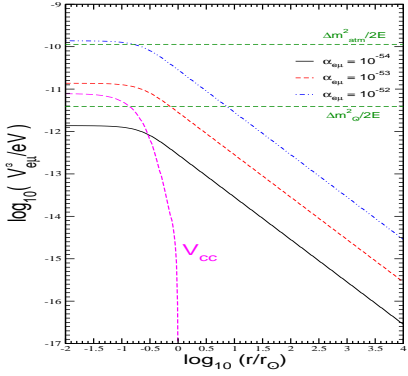,height=5cm,width=6cm,angle=0}}
\caption{Comparison of the MSW potential $V_{cc}$ 
and the LR potential $V_{e\mu}^\odot$ due to the solar electrons.
The $(\dmsq/2E)$ values
corresponding to $E=10$ MeV are also shown.
\label{fig:vemu}
 }
\end{figure}

The behavior of $V_{e\beta}^\odot$ from the solar core 
all the way to the Earth ($r \approx 215~ r_\odot$)
and beyond is shown in Fig.~\ref{fig:vemu}.
It is seen that 
$V_{e\beta}^\odot$ dominates over the MSW potential $V_{CC}$
inside the Sun for $\alpha \gsim 10^{-53}$.
Moreover, it does not abruptly go to zero outside the Sun 
like $V_{CC}$, but decreases inversely with $r$.
Its value at the Earth surface is an order of magnitude
larger than the potential there due to the electrons
in the Earth, so one can neglect the latter.

When $R_{LR} \gsim R_{\rm gal}$, our distance from
the galactic center, the collective potential due to
all the electrons in the galaxy may become significant. 
We denote the galactic contribution to the potential $V_{e\beta}$ as
\beq
V_{e\beta}^{\rm gal} = b ~ \alpha_{e\beta}~ 
N^0_{e, {\rm gal}}/R^0_{\rm gal} \; ,
\label{b-def}
\eeq
where $N^0_{e, {\rm gal}} \equiv 10^{12} N_e^\odot$
and $R^0_{\rm gal} \equiv 10$ kpc.
The parameter $b$ takes care of our ignorance about the distribution of
the baryonic mass in our galaxy. With $R_{LR} \gsim R_{\rm gal}$,
we expect $0.05 < b <1$. The value of $b$ may be smaller 
if $R_{LR}$ is smaller,
$b=0$ is equivalent to  $R_{LR} \ll R_{\rm gal}$.
The screening effects\cite{screening} 
are negligible over the scale of $R_{\rm gal}$.

\section{Masses, Mixings and Resonances of Solar Neutrinos}
\label{analytic}

The LR potential gives unequal contributions to all three flavors 
simultaneously  unlike
in case of the charged current potential.
As a result, the inclusion of three generations in the
solar analysis becomes necessary.

The appropriate Hamiltonian 
in the $L_e-L_\mu$ case 
describing the neutrino propagation 
can be written in the flavor basis as 
\beq 
\label{hf}
H_f = R ~ H_0 ~ R^T +V \;,
\eeq
with $V = {\rm Diag}(V_{cc}+V_{e\mu}, -V_{e\mu}, 0)$.
Here $R \equiv R_{23}(\tb) R_{13}(\tc) R_{12}(\ta)$,
we have assumed that no CP violation enters the picture.
One can take 
$H_0 = {\rm Diag}(0, \Delta_{21}, \Delta_{32})$
with $\Delta_{ij} \equiv \dmsq_{ij}/(2E)$.
The antineutrino propagation is obtained by
the replacement 
$V\rightarrow -V$.

\begin{figure}
\centerline{\psfig{figure=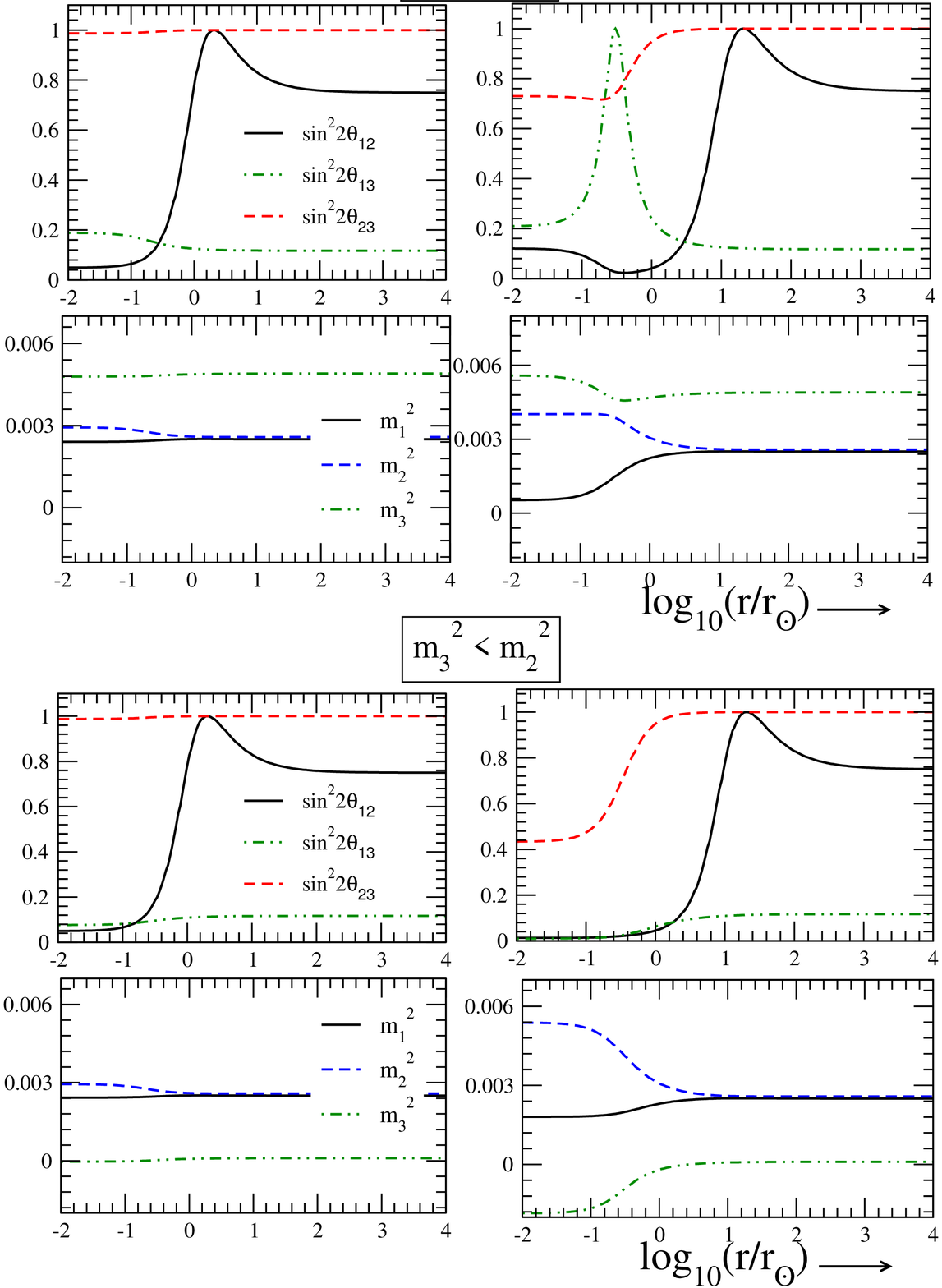,height=7cm,width=6cm,angle=0}}
\caption{The angles and $m_i^2$ values in matter for solar 
neutrinos for $E=10$ MeV, 
in the case $R_{LR} \ll R_{\rm gal}$. The $m_i^2$
values are correct up to an additive constant, so that only
their relative values have a physical significance.
\label{thetas}
}
\end{figure}

The Hamiltonian (\ref{hf}) can be analytically diagonalized\cite{aaa} 
by keeping terms linear in the small parameters 
$x \equiv \Dsol/\Datm \approx 0.03$
and $\sin \theta_{13}<0.2$, except in a narrow region around
$y_{e\mu}\equiv V_{e\mu}/\Delta_{32} \approx 2/3$.
The exact numerical values for mixing angles and $m_i^2$ 
for different values of $\alpha$ for normal as well as
inverted hierarchy are shown in Fig.~\ref{thetas}.

\subsection{For $\alpha \lsim 10^{-52}$ }
\label{smallalpha}

Both $\theta_{23}$ as well as $\theta_{13}$ get only small 
corrections. 
In the limit $\theta_{13m} \to 0$, the third mass eigenstate 
decouples and the scenario reduces to 2$\nu$ mixing.
However, the effective matter potential is\cite{aaa} 
\beq
{\cal V}_{12} \approx V_{cc} + V_{e\mu} (1+\cos^2 \theta_{23m}) ~, 
\label{v12}
\eeq
and not $V_{cc}+ 2 V_{e\mu}$ as would have been taken in a naive
2-generation analysis.

The MSW resonance takes place with the modified potential
${\cal V}_{12}$.
For $\alpha \gsim 10^{-53}$, the ${\cal V}_{12}$ 
contribution dominates over $V_{cc}$ and
the resonance is shifted outside the Sun where its 
behavior is solely determined by the LR potential.

If $P_L (E)$ is the jump probability at the $\nu_{1m}$-$\nu_{2m}$
resonance, the net survival probability of $\nu_e$ is
\barr
P_{ee}(E) & = &
(1-P_L)~ c^2_{13P}~ c^2_{12P}~ 
c^2 _{13E} ~c^2_{12E} 
 \nonumber \\
& + & P_L~ c^2 _{13P}~ s^2_{12P}~ 
c^2 _{13E} ~c^2_{12E} 
 \nonumber \\
& + & (1 - P_L)~ c^2 _{13P}~ s^2 _{12P} 
~c^2_{13E} ~s^2_{12E} 
\nonumber \\
& + & P_L~ c^2 _{13P} ~c^2 _{12P}~ 
c^2_{13E} ~s^2_{12E} 
\nonumber \\
& + & s^2 _{13P}~ s^2_{13E} \; .
\label{pee}
\earr
Here $\theta_{ijP}$ and $\theta_{ijE}$ 
are the values of $\theta_{ijm}$ at
the neutrino production point and at the Earth respectively.

\subsection{For $\alpha {\gsim} 10^{-52}$}
\label{largealpha}

In this range of $\alpha$,
the $\nu_{1m}$-$\nu_{2m}$ resonance
is always outside the Sun and adiabatic.
In addition 
the angle $\theta_{13m}$ gets resonantly enhanced when
$y_{e\mu} \approx 2/3$ (normal hierarchy).
The $\theta_{13m}$ enhancement corresponds to the 
$\nu_{2m}$-$\nu_{3m}$ level crossing, with an effective
potential\cite{aaa}
\beq
{\cal V}_{23} = V_{cc} + V_{e\mu} (1+\sin^2 \theta_{23m})~.
\label{v23}
\eeq
The net survival probability of $\nu_e$ is
\barr
P_{ee}(E) & = &
c^2 _{13P} ~c^2_{12P}~ 
c^2 _{13E}~ c^2_{12E} 
 \nonumber \\
& + & (1 - P_H)~ c^2 _{13P}~ s^2 _{12P} 
~c^2_{13E}~ s^2_{12E} 
\nonumber \\
& + & P_H~ s^2 _{13P} 
~c^2 _{13E} ~s^2_{12E} 
\nonumber \\
& + & (1 - P_H) ~s^2 _{13P} ~s^2_{13E} 
\nonumber \\
& + & P_H ~c^2 _{13P}~ s^2 _{12P}~
s^2 _{13E} ~,
\label{pee-3nu}
\earr
where $P_H(E)$ is the probability that $\nu_{2m}$ and $\nu_{3m}$
convert into each other after traversing through this resonance. 
	
In general $P_H \approx 0$ at high values of $\theta_{13}$.
For $\theta_{13} \lsim 0.1^\circ$, the value of $P_H$
becomes significant. 
In the range where $0.1 < P_H < 0.9$ (the semi-adiabatic range), 
$P_H$ is also highly energy dependent.

\section{Constraints from solar neutrinos and KamLAND}
\label{numerical}

We perform a global fit to the data from solar experiments
and KamLAND, using the $\ch$ minimization technique
with covariance approach for the errors.\cite{ana2g_lisicorr}

\begin{figure}[t]
\centerline{\psfig{figure=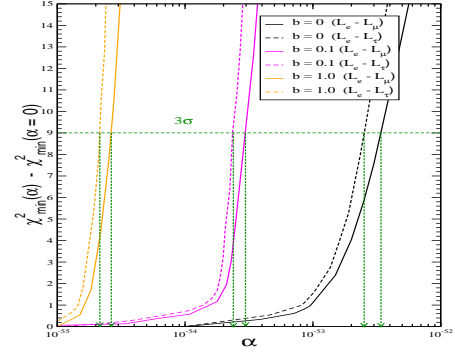,height=5cm,width=6cm,angle=0}}
\caption{  $\Delta \chi^2$ values and limits for the $L_e-L_\mu$ 
as well as $L_e - L_\tau$ symmetry,
with $\theta_{13}=0^\circ$.The case $R_{LR} \ll R_{\rm gal}$
is represented by $b=0$ and higher $b$ values
correspond to larger contributions from galactic
electrons (see Sec.~\ref{galaxy}).
\label{chisq-emu}}
\end{figure}
  
The best fit values for the solar parameters are always observed
to lie in the LMA range\cite{lma} with vanishing $\alpha_{e\mu}$ giving
the best fit.
For $\alpha < 10^{-52}$, the value of $\chi^2$ is minimum for
$\theta_{13}=0^\circ$.
When $\alpha > 10^{-52}$, a strong energy dependence in
the survival probability is introduced for $\theta < 0.1^\circ$
through $P_H(E)$,
so that the $\chi^2$ values for extremely low $\theta_{13}$
values become large. 
However, the region $\alpha > 10^{-52}$
turns out to be  excluded to more than 3$\sigma$.
Therefore we quote the most conservative
upper bounds on $\alpha$, by
taking $\theta_{13} = 0^\circ$. 
These limits are shown in Fig.~\ref{chisq-emu}:
the $3\sigma$ limit corresponding to the one-parameter fit 
in the $L_e-L_\mu$ case is
$\alpha_{e\mu} < 3.4 \times 10^{-53}$.
The corresponding $L_e-L_\tau$ limit is
$\alpha_{e\tau} < 2.5 \times 10^{-53}$.
The bounds are independent of the neutrino mass hierarchy.

\section{LR potential from the galaxy}
\label{galaxy}

\begin{figure}
\centerline{\psfig{figure=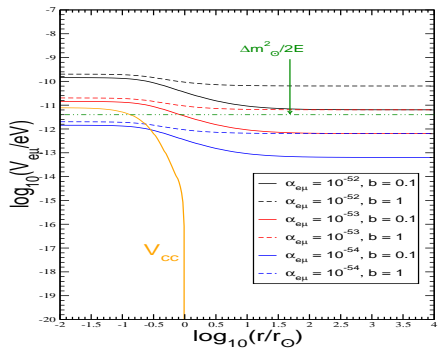,height=5cm,width=6cm,angle=0}}
\caption{The net potential
$V_{e\mu} \equiv V_{e\mu}^\odot + 
V_{e\mu}^{\rm gal}$ for various values of $b$ and $\alpha$.
\label{vemu-net}}
\end{figure}

When $R_{LR} \gg R_{\rm gal}$, the net potential
$V_{e\mu}  \equiv V_{e\mu}^\odot + 
V_{e\mu}^{\rm gal}$ is shown in Fig.~\ref{vemu-net}.
For $V_{e\mu}^{\rm gal} \gg \dmsq_\odot /(2E)$,
there is no MSW resonance that is essential for a good fit 
to the solar neutrino data. 
Even for lower values of $b$ and $\alpha$,
if $V_{e\mu}^{\rm gal}$ dominates over $V_{CC}$ 
at the MSW resonance, 
the resonance tends to be adiabatic 
even for low energies, so the radiochemical data 
disfavors the solution.

The $\Delta \chi^2$ values as a function of $\alpha$ for
different $b$ values are shown in Fig.~\ref{chisq-emu}.
The 3$\sigma$ constraints for the $L_e-L_\mu$ case are
$\alpha_{e\mu}  <  2.9 \times 10^{-54} ~(b=0.1)$ and
$\alpha_{e\mu}  <  2.6  \times 10^{-55} ~(b=1)$. 
For $L_e-L_\tau$, the constraints are
$\alpha_{e\tau}  <  2.3 \times 10^{-54} ~(b=0.1)$ and
$\alpha_{e\tau}  <  2.1 \times 10^{-55} ~(b=1)$. 
We expect $b<1$ even with generous estimates.
The most conservative constraints are clearly with
$b=0$, as calculated in Sec.~\ref{numerical}.

\section{Concluding remarks}
\label{summary}

The long range forces mediated by $L_e-L_{\mu,\tau}$ vector gauge
bosons can give rise to non-trivial 3-$\nu$ mixing effects
inside the Sun, and affect the MSW resonance picture.
The angle $\theta_{13}$ may even get resonantly amplified
if $\alpha \gsim 10^{-52}$ for normal hierarchy.
These effects allow one to put constraints on the coupling
of the LR forces from the solar and KamLAND data.
The bounds obtained are orders of magnitude
better than those obtained earlier from the 
gravity experiments and atmospheric neutrino data.
If the range of the forces is larger than $R_{\rm gal}$, the
bounds become still stronger by upto two orders of magnitude.

A recent paper\cite{garcia} also has given comparable 
bounds on the LR couplings. 
However, they assume one mass scale dominance,
neglecting the effect of the third neutrino altogether.
Moreover, they have not taken
the galactic contribution into account even when
the range of the force is more than $R_{\rm gal}$.

\section*{Acknowledgments}

AD would like to thank B. Dasgupta and G. Raffelt for 
useful discussions.
The work of AB and AD is partly supported through the 
MPP-TIFR Partner Group project.


\end{document}